\documentclass{article}

\usepackage{microtype}
\usepackage{graphicx}
\usepackage{epstopdf}
\usepackage{subfigure}
\usepackage{amsmath}
\usepackage{booktabs} 

\usepackage{hyperref}

\usepackage[accepted]{icml2020}

\icmltitlerunning{Generative Adversarial Networks with Conditional Neural Movement Primitives for An Interactive Generative Drawing Tool}

\begin{document}

\twocolumn[
\icmltitle{Generative Adversarial Networks with Conditional Neural Movement Primitives for An Interactive Generative Drawing Tool}



\icmlsetsymbol{equal}{*}

\begin{icmlauthorlist}
\icmlauthor{Suzan Ece Ada}{equal,boun}
\icmlauthor{M. Yunus Seker}{equal,boun}
\end{icmlauthorlist}

\icmlaffiliation{boun}{Computer Engineering Department, Bogazici University, Istanbul, Turkey}

\icmlcorrespondingauthor{Suzan Ece Ada}{ece.ada@boun.edu.tr}
\icmlcorrespondingauthor{M. Yunus Seker}{yunus.seker1@boun.edu.tr}

\icmlkeywords{Machine Learning, ICML}

\vskip 0.3in
]



\printAffiliationsAndNotice{\icmlEqualContribution} 

\begin{abstract}
Sketches are abstract representations of visual perception and visuospatial construction. In this work, we proposed a new framework, Generative Adversarial Networks with Conditional Neural Movement Primitives (GAN-CNMP), that incorporates a novel adversarial loss on CNMP to increase sketch smoothness and consistency. Through the experiments, we show that our model can be trained with few unlabeled samples, can construct distributions automatically in the latent space, and produces better results than the base model in terms of shape consistency and smoothness.
\end{abstract}

\section{Introduction}
\label{intro}


Sketches are abstract representations of visual perception and visuospatial construction. Therewith, sketching is to communicate visual mental imagery in art and engineering extensively. Human-AI collaboration in sketch generation can aid humans in conveying their designs through completion, correction, and generation. Since sketch data are abstract representations of mental imagery, sketch generation, and correction tasks provide exceptional benchmarks for learning techniques. 

One learning technique used for sketch generation and correction tasks is generative modeling. Existing methods use Recurrent Neural Networks(RNN) \cite{RNN}, Generative Adversarial Networks \cite{goodfellow2014generative} (GANs), and Autoencoders \cite{autoenc}, and Reinforcement Learning (RL) to train sequential data. Yet, RNN based architectures require large datasets for training and fall short in processing data independently. This dependency often results in error accumulation. To overcome these challenges, we use Conditional Neural Movement Primitives \cite{cnmp} (CNMP) architecture that can fit few data and process/generate data independently. We integrate a discriminator to the CNMP architecture to increase smoothness and decrease generated shape inconsistency. These models can represent multi-class data, generate the remaining of the unfinished samples, and correct non-perfect samples by using few samples from each class. 

\begin{figure}[t]
\vskip 0.1in
\begin{center}
\centerline{\includegraphics[width=0.8\columnwidth]{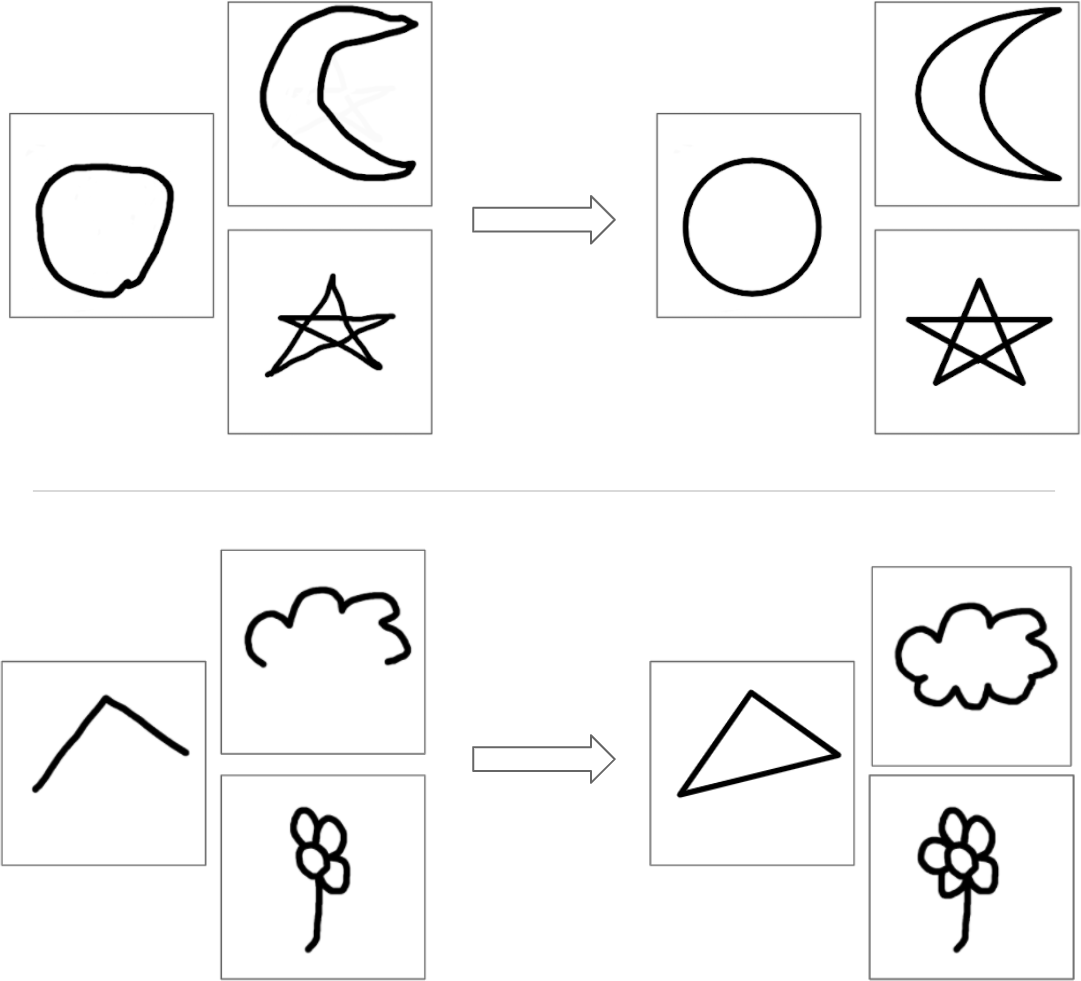}}
\caption{Auto-correct (first row) and auto-complete (second row) features}
\label{autocorrect}
\end{center}
\vskip -0.4in
\end{figure}

Our aim in this work is to generate auto-corrected and auto-completed shapes/drawings from non-perfect human cursor drawings as in Figure \ref{autocorrect}. More specifically, the generator will remove the noise in the human cursor drawings and generate a smoother version of the input sketch in the auto-correction task. Likewise, the remaining part of the unfinished sketch will be generated by our generator model in the auto-completion task. Our model will infer the correct shape class and the size of the sketch merely from the cursor trajectory observations. 

In this paper, we incorporate a novel adversarial loss in CNMP to increase sketch smoothness and consistency. Furthermore, the discriminator encourages the generation of sketches that are more similar to training data. We regard sketch generation as a trajectory generation task where each sketch is analogous to a trajectory that consists of time-steps and corresponding 2D values. We use CNMP as the conditional generator and a feed-forward network as the discriminator. The objective of the discriminator is to discriminate the true sketch given the generated or the true sketch. We compare GAN-CNMP with CNMP and experimentally show that the sketches generated by GAN-CNMP are closer to ground truth sketches.




\section{Related Work}
\label{relatedwork}
Substantial progress has been made in Human-AI collaboration tasks through the efficacious application of generative modeling techniques. 

\paragraph{Sketch Generation}

Sketches can be represented in various ways including trajectories, pixels and pen state information. This representation diversity allows the usage of various different architectures like RNNs \cite{ha2017neural}, GAN\cite{goodfellow2014generative,v2019teaching}, RL\cite{zhou2018learning,ganin2018synthesizing} , VAE \cite{v2019teaching}, and transformers\cite{ribeiro2020sketchformer,wieluch2020strokecoder} for sketch generation. 

One way of depicting sequential sketch information data is vectorizing point location information. QuickDraw dataset constructed in Sketch-RNN \cite{ha2017neural}, consists of 75K sketches where each sketch is represented by 5D-vector points. A point is a concatenation of 2D point offset distance from the previous point and the pen state information. The discrete pen state information contains one-hot-encoded information of 3 different binary states: the pen will continue to touch the paper, the pen will not touch the paper and the drawing has stopped. Sketch-RNN is a sequence-to-sequence VAE architecture that consists of a bi-directional RNN encoder that encodes sequential sketch data to a latent vector and an auto-regressive RNN decoder that generates the next step in the sequence. Because the encoded latent vector and the sampled output are stochastic, the randomness of the generated sketches is versatile. This flexibility allows diverse ways of completing existing sketches and conditional reconstruction. 
 
Similar to Sketch-RNN, in \cite{v2019teaching} the same vectorized data representation is used to generate sketches using different GAN architectures namely SkeGAN and VAEske-GAN. SkeGAN uses a policy network as the generator and a bidirectional LSTM\cite{lstm} as the discriminator. The stochastic policy network is optimized using the reward signal from the discriminator that outputs the probability that the generated state sequence is fake or not. In VAEske-GAN, a bidirectional LSTM is used to encode the sketches as a latent vector. The discriminator decides whether the sketches generated by the decoder LSTM are fake or not. They conduct a comparative analysis on SkeGAN and VAE-GAN using a metric named Ske-Score \cite{v2019teaching} which aims to decrease the scribble effect mentioned in \cite{ha2017neural}. Ske-Score of a sketch is found by dividing the number of times the pen does not touch the paper by the number of times it touches the paper. However, to obtain the Ske-Score, the dataset requires the pen state information. The generator components of SketchRNN, SkeGAN, and VAEske-GAN are trained using sketches from a particular class to complete an unfinished sketch from the same category. 

Modeling the sequential sketch data as an RL problem has been gaining interest in the sketch generation domain. In \cite{zhou2018learning}, canvas states and pen actions were used to to accomplish the task in the RL framework. The actions are predicted based on the local and global canvas states extracted by two respective CNNs. Q-function is used as a reward signal to reproduce the real sketches. Combined with supervised imitation learning, Doodle-SDQ \cite{zhou2018learning} can successfully extrapolate to classes not seen during training. 

There has been an increasing trend in replacing the LSTMs with transformer-based architectures in the computer vision domain. In particular, Sketchformer\cite{ribeiro2020sketchformer} and StrokeCoder\cite{wieluch2020strokecoder} are recent methods that utilizes transformers for sketch generation. In Sketchformer, a 5D-vector representation of the QuickDraw dataset is used to train the model. The original transformer architecture in \cite{DBLP:journals/corr/VaswaniSPUJGKP17} is modified by adding a sketch embedding layer in the bottleneck and increasing both the number of multi-head attention blocks and the feed-forward dimension. They obtained a 6 percent increase in performance compared to the Sketch-RNN (an LSTM based autoencoder technique). In Strokecoder, first, data augmentation is applied by translation, rotation, scaling, and mirroring to the stroke-based images before feeding them to the transformer encoder.


\begin{figure}[]
\begin{center}
\centerline{\includegraphics[width=\columnwidth]{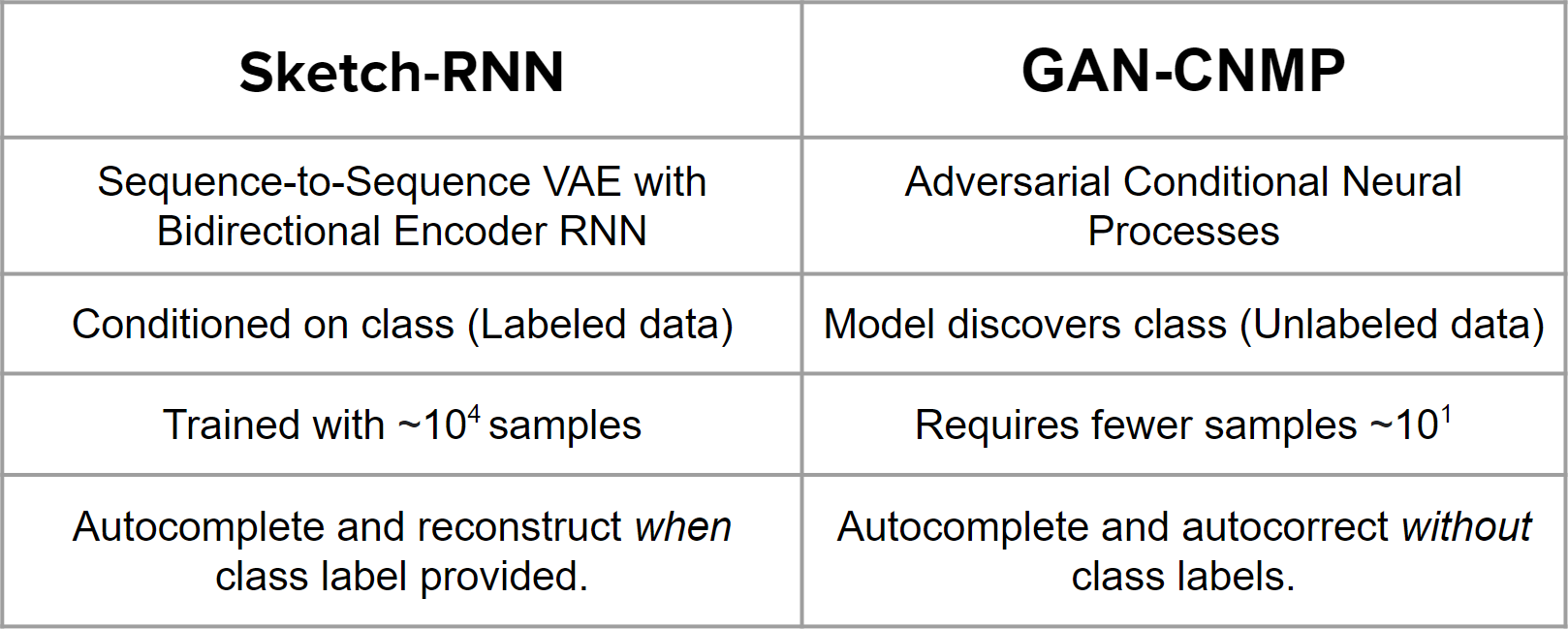}}
\caption{Comparison of our method with Sketch-RNN}
\label{sketchrnn}
\end{center}
\vskip -0.35in
\end{figure}



\paragraph{Conditional Neural Processes}

CNP and CNMP are generative models that integrate stochastic processes in neural networks via latent variable conditioning. In both CNP and CNMP, randomly sampled data points are fed separately to identical encoders to form an average prior latent representation. This representation is then concatenated with desired points for generation and fed to the query network. Inspired by the fruitful results obtained in pixel-wise image completion task on CelebA dataset using CNP and robotic movement generation task using CNMP we make use of the similar architecture in the sketch generation task.

\section{Method}
\label{method}

In this section, first, background information about the base model that is used in our framework will be provided. After that, our proposed method will be explained in detail.

\subsection{Background: Conditional Neural Processes}

Conditional Neural Process (CNP) is a new type of neural network architecture that works based on Bayesian Inference principles. CNP transfers the advantages of Gaussian Processes \cite{GP} (GP) to deep learning and improves the weaknesses of its predecessor by using the benefits of neural network architectures. Probabilistic approximators generally become handy when the training data consists of distributions, such as Gaussian's, that can be represented by statistical methods. The classical Gaussian Processes have some drawbacks when it comes to using them in practice, however, CNP proposes solutions to all of these drawbacks by using the advantages of neural network architectures. 

The first drawback of GP is having to define a prior distribution. The posterior distribution is fit to the training data by assuming that the prior distribution fully covers the characteristics of the actual training distribution. However, having to define a prior distribution can be troublesome in most cases because of the complex and high-dimensional data characteristics. CNP proposes a solution to that by introducing an encoder structure that samples observations from the data, and extracts prior knowledge directly from it without needing any prior definition. The encoder transfers the input space to a latent space to construct meaningful representations of the actual distribution. These representations are then used to predict conditional distributions for any target inputs in the latter stages of the training.

The second drawback of GP is that the complexity of the method in the test time is considerably high. A classical GP has $O(n+m)^3$ complexity in the test time where n is the number of observations, and m is the number of target inputs. The numbers and the cost of the algorithm can easily go vague when the n and m numbers are starting to increase according to the use cases of the method. CNP proposes a solution to this problem and reduces the complexity to $O(n+m)$ by fixing the dimensionality of the representation space. This trade-off provides the framework flexibility and scalability.

The third drawback of the GP is the lack of capacity to handle high-dimensional data. The complexity and the structure of the classical method cause the usage of high-dimensional data such as images to become difficult. CNP offers a solution to this problem by using the benefits of neural networks and layers such as Convolutional Neural Networks. Convolutional layers drastically reduce the complexity of high-dimensional data, especially with GPUs.

\begin{figure}[t]
\vskip 0.2in
\begin{center}
\centerline{\includegraphics[width=\columnwidth]{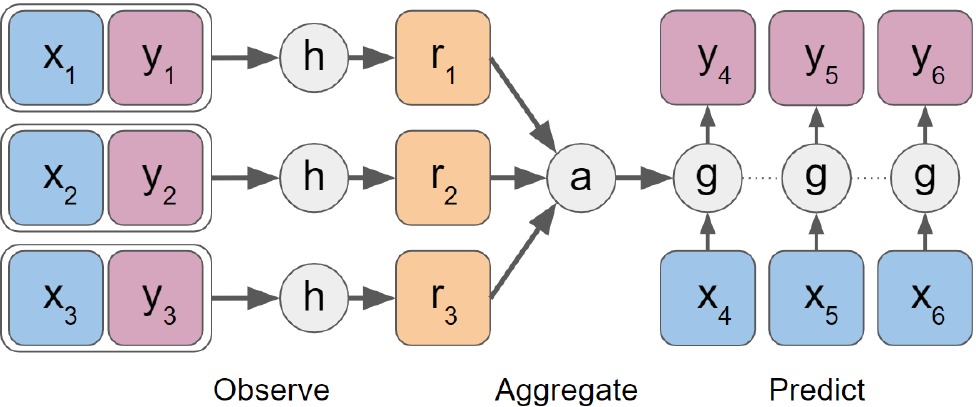}}
\caption{Structure of the CNP adapted from \cite{cnp}}
\label{cnp}
\end{center}
\vskip -0.2in
\end{figure}

\begin{figure*}[ht]
\vskip 0.2in
\begin{center}
\centerline{\includegraphics[width=\linewidth]{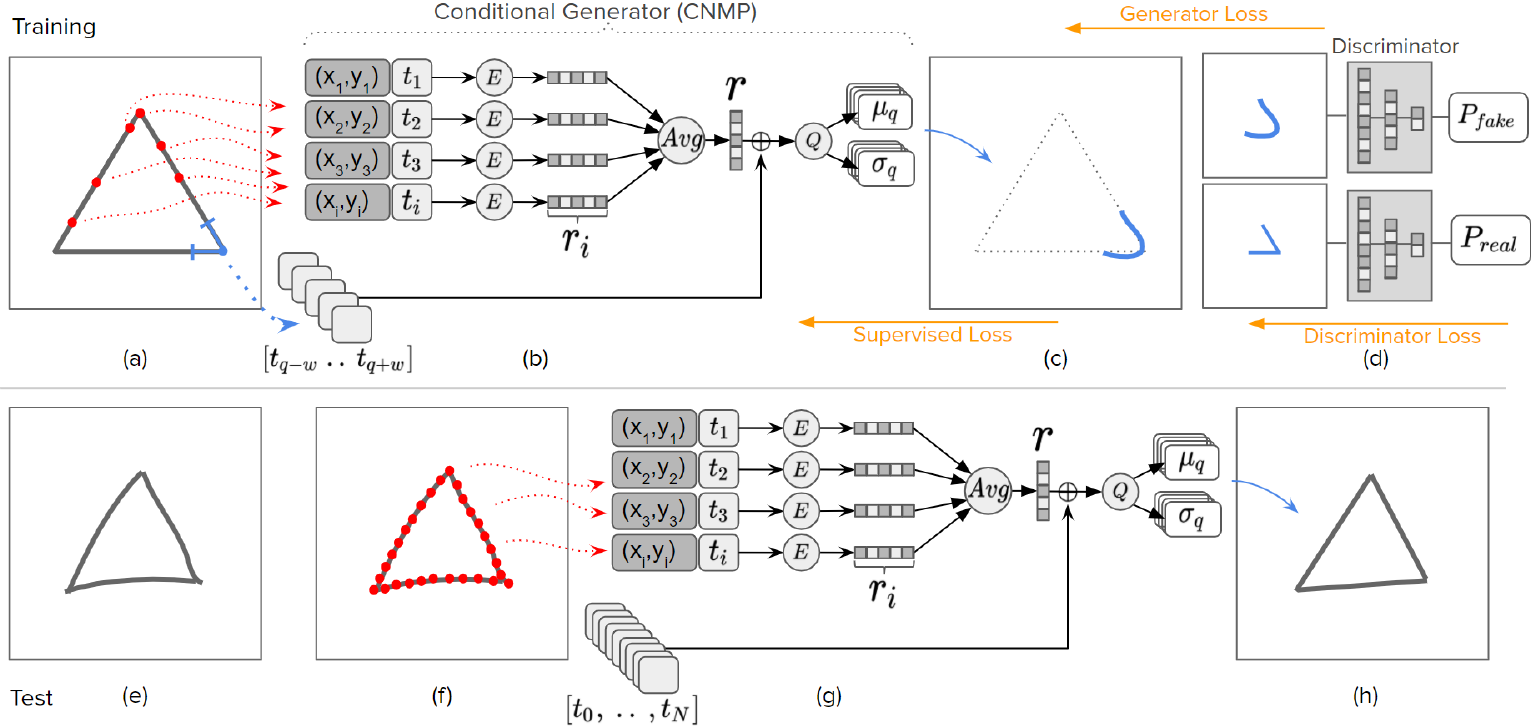}}
\caption{General framework of GAN-CNMP. First row shows the training process. Second row shows how the drawing are regenerated using the trained model.}
\label{main}
\end{center}
\vskip -0.2in
\end{figure*}

Figure \ref{cnp} shows the general structure of the CNP model that consists of three parts. (Observe) Assuming there is a function family $F$ which $f_i(x)=y$ and $f_i \in F$, a random number of input and output observations ($x,y$) are sampled from $F$. Then, these observations are passed through a parameter sharing encoder network (h) to construct their corresponding latent space representations ($r_i$). Using a parameter sharing encoder network allows CNP to process the observations independently from each other as well as scaling the network to multiple and changing input size which is a dynamic property provided in the Gaussian Processes. (Aggregate) To handle multi-inputs, all of the representations are aggregated into one general representation by using an averaging layer (a). This general representation holds the characteristics of all representations, thus, all observations in the first place. This knowledge is used for predicting conditional distributions over other target inputs. (Predict) A decoder network uses the concatenation of the general representation $r$ and target input $x_i$ and predicts a conditional distribution ($\mu_i,\sigma_i$). The general CNP model can be expressed as: \[\mu_q,\sigma_q = g_\theta\left( x_q \oplus \frac{\sum_i^{n} h_\phi( (x_i,y_i))}{n}\right)\] where $\{(x_i,y_i)\}$ are observation pairs sampled from data , $h_\phi$ is the encoder network with parameters $\phi$, $x_q$ is the target input, $\oplus$ is the concatenation operator, $g_\theta$ is the query network with parameters $\theta$, and $(\mu_q,\sigma_q)$ are the outputs the Gaussian distribution parameters. The whole model is trained stochastically and end-to-end. Network parameters ($\theta$ and $\phi$) are optimized according to the following loss function which will be called reconstruction loss later:

\begin{equation} \label{eq:lossFunction}
    \mathcal{L}(\theta,\phi) = -\log P(y_q \mid \mu_q,\mathop{\textrm{softplus}}(\sigma_q))
\end{equation}
where $\mu_q$ and $\sigma_q$ are the outputs of the CNP, $y_q$ is the ground-truth output of the $x_q$ for the sampled function $f_i \in F$, and $P$ are the Gaussian probability function that returns the conditional probability of the $y_q$ for the given mean and variance parameters.

\paragraph{Conditional Neural Movement Primitives (CNMP)}
CNMP\cite{cnmp} is proposed as robotic learning from demonstration framework that uses CNP as their base model. Despite being a relatively different research area which is out of the scope of this paper, their framework actually has a really similar approach to use the data and the base model with our proposed method. In their paper, the authors showed that CNP can be modified to be used to learn trajectory distributions which are functions from time to Cartesian coordinates of the robot manipulator in three-dimensional space. They also showed that training CNPs require fewer training samples and perform better in learning unlabeled data compared to other classical sequential learning algorithms such as LSTMs\cite{lstm}. In our paper, we will be working on drawing trajectories which are functions from time to two-dimensional coordinates of the cursor on the screen. Because of the similarities in the use cases, and the advantages of the CNMP on learning unlabeled data with few samples, we will be using CNMP as our conditional generator for the drawings, which will be explained in the next subsection in details.

\subsection{Proposed Method: GAN-CNMP}
Although CNMPs have an advantage over the classical methods, every timestep is predicted independent from each other, thus, there are no mathematical guarantees that ensure the smoothness and consistency between two sequential target inputs. In this paper, we propose a new framework as a solution to these problems by using a generative model architecture on top of the base CNMP model.

We define our train set as a collection of drawing trajectories, $D_i \in D_{i=0}^N$, where $D$ is the training set collection, and $N$ is the total number of drawings in the training set. Each drawing is represented as a function from time to 2D cursor position on the screen and discretized to $M$ samples for each drawing in order to make the training process smoother. A drawing is defined as $D_i = \{(t_j,(x_j,y_j))\}_{j=0}^{M}$, where $t_j$ is the $j$th timestep in the trajectory, and $(x_j,y_j)$ are the 2D positions of the cursor on the screen at time $t_j$.

Figure \ref{main} shows the general structure of our framework where the first and second rows explain the training and test phases respectively. The training process consists of 4 steps which will be explained below. The framework has two parts which are conditional generator (CNMP) and discriminator. 

In the first part, a drawing is generated by the conditional generator, and reconstruction loss is calculated. At the start of each training iteration, a random drawing trajectory $D_i$ is selected from the training set $D$ (Fig. \ref{main}a). Then, from that drawing, a random number of maximum $obs_{max}$ time and cursor position pairs are sampled as observations, $O={(t_k,(x_k,y_k))}_{k=0}^{obs_{max}}$ and $(t_k,(x_k,y_k)) \in D_i$. Note that $obs_{max}$ is a hyper-parameter that changes in every training iteration in order to make the framework robust to the changing number of observations at the test time. Besides the observations, a random target time window is also sampled from the same drawing, $T = [t_{q-w}, .. , t_q , .. , t_{q+w}]$, where $t_{q}$ is the randomly selected timestep, and $w$ is the size of the target time window which is another hyper-parameter in the system. 

The purpose of the conditional generator is to generate a drawing on the target time steps $T$ based on the given conditional information of the observation set $O$ (Fig. \ref{main}b). In our study, we used CNMP architecture as our conditional generator. The observations are passed through a parameter sharing encoder and then they are merged into one general representation by using an averaging layer. After that, this representation is replicated and concatenated with the elements of $T$, and passed through a decoder network in order to predict a conditional distribution for each target timestep.

After the generator produces a drawing for the given target points based on the observations (Fig. \ref{main}c), the output of the generator is compared with the ground-truth values of the drawing in order to calculate the reconstruction loss of our system, $\mathcal{L}_{rec}$ which is the first loss function used in the framework. Note that Eq. \ref{eq:lossFunction} and Eq. \ref{eq:recFunction} are nearly the same loss function with the only difference that the reconstruction loss used in our system is averaged over the multiple numbers of target time points in the $T$.
\begin{equation} \label{eq:recFunction}
    \mathcal{L}_{rec} = -\sum_{k}^{T}\log P(x_k,y_k \mid \mu_k,\mathop{\textrm{softplus}}(\sigma_k))
\end{equation}

In the second part of the training, which is the main contribution of our paper, the generated drawing and its corresponding ground-truth trajectory are given to a discriminator, called $D_{c}$, in order to distinguish the real drawing from the fake (generated) one. $D_{c}$ is a multi-layer perceptron that consists of fully connected layers, and outputs a probability between $[0,1]$ that represents the belief of the discriminator on the input to be real (ground-truth). These outputs are used to calculate the two additional losses that we proposed in our system which are generator $\mathcal{L}_{G}$ and discriminator loss $\mathcal{L}_{D}$:

\begin{equation}
\label{eq:ganLoss}
\begin{aligned}
&\mathcal{L}_{G} = -\log(D_c(G(O,T)))\\        
&\mathcal{L}_{D} = -\log(1-D_c(G(O,T))) - \log(D_c((x,y)_T))
\end{aligned}
\end{equation}

where $D_c$ is the discriminator, $G$ is the conditional generator which is the CNMP model, and $(x,y)_T$ is the ground-truth 2D cursor positions of the selected drawing on the target time window. The losses are calculated according to basic Generative Adversarial Network architectures and adapted to our framework. Finally, the general loss of our system is calculated as:

\begin{equation}
\label{eq:mainLoss}
\begin{aligned}
\mathcal{L} = \mathcal{L}_{rec} + \mathcal{L}_D + \mathcal{L}_G
\end{aligned}
\end{equation}

The second row of the Figure \ref{main} shows the test phase of our framework which consists of only the generator part. In the test time, users draw a shape to the screen in order to give input to our framework (Fig. \ref{main}e). This cursor trajectory is recorded with the related time information and sampled into $M$ timestep and cursor position to make it similar to the trajectories sampled in the training set (Fig. \ref{main}f). After that, all of the sampled points are given to our model as observations $O$. After that, the target time samples $T$ are selected as all timesteps since our aim in test time is to regenerate the whole shape from the start to the end. These observations and target timesteps are given to the generator (CNMP) and our framework regenerates a drawing according to the observations taken from the drawing made by a human at the start (Fig. \ref{main}g). Finally, a drawing that expresses the knowledge learned during the training phase while preserving the characteristics of the drawing made in the test time is generated (Fig. \ref{main}h).

\section{Dataset}
\label{sec:dataset}
\begin{figure}[h]
\vskip 0.2in
\begin{center}
\centerline{\includegraphics[width=\linewidth]{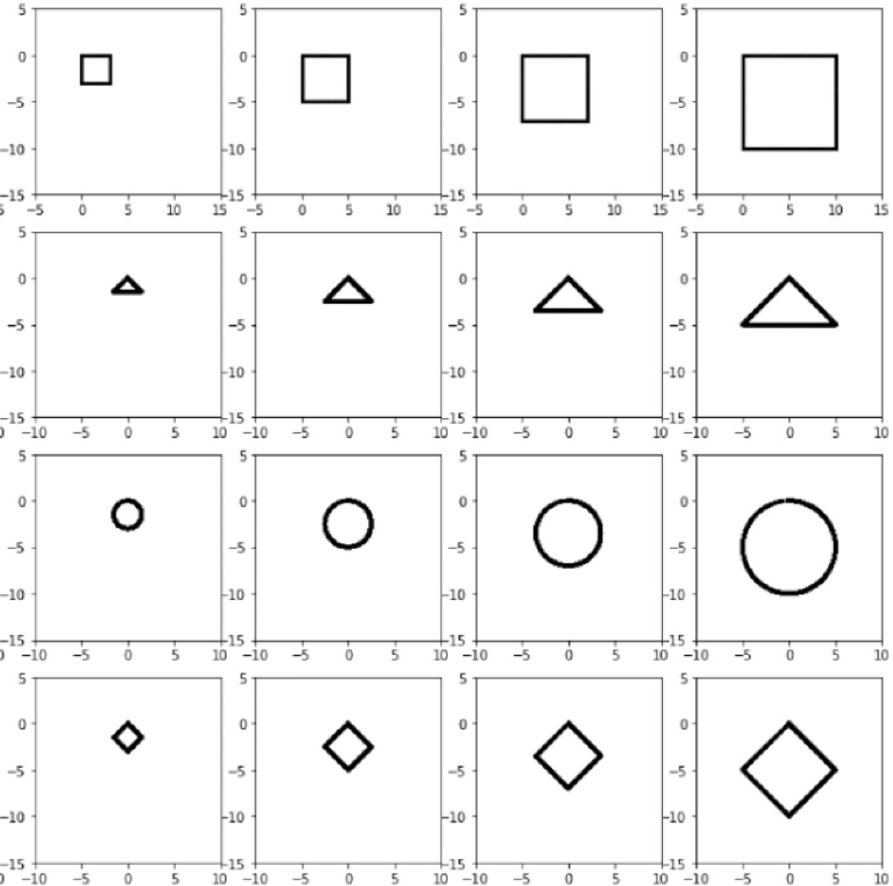}}
\caption{Different size of examples from the training set for 4 shape types.}
\label{dataset}
\end{center}
\vskip -0.2in
\end{figure}

Using a small shape dataset that consists of 40 human cursor drawings, we experimentally and theoretically prove that our method can fit small datasets. We construct the dataset by generating 10 different sizes for each shape class by using their mathematical formulas: square, triangle, circle, and diamond shown in Figure \ref{dataset}. For each drawing, we record two-dimensional position information for 100 timesteps.

\begin{figure*}[h]
\vskip 0.2in
\begin{center}
\centerline{\includegraphics[width=\linewidth]{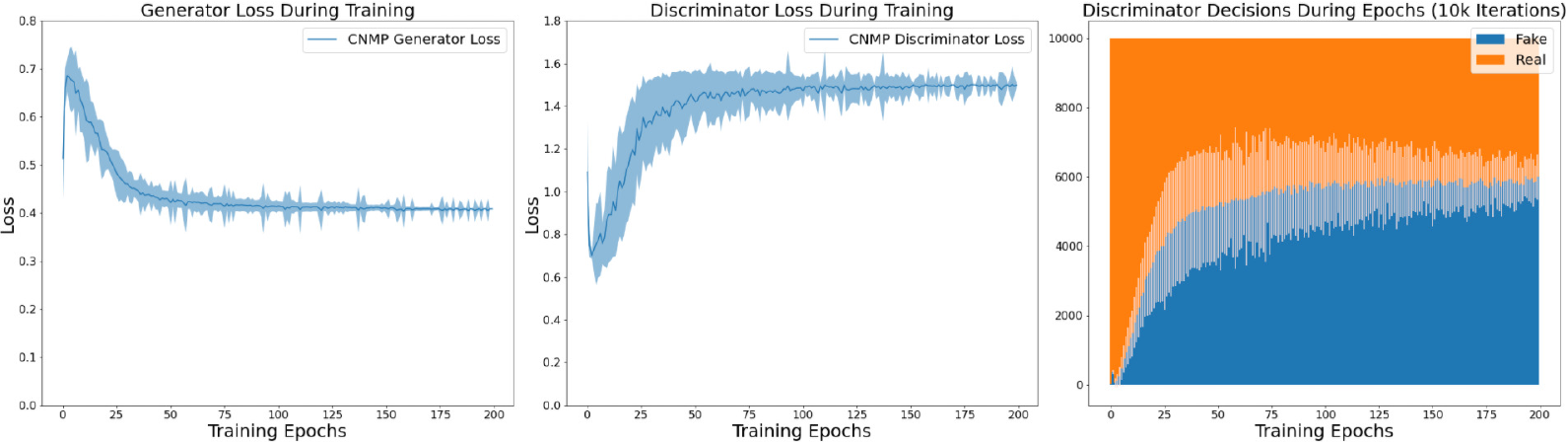}}
\caption{Generator (left) and Discriminator (middle) losses during the training. Discriminator decisions are shown on the right.}
\label{training}
\end{center}
\vskip -0.2in
\end{figure*}

\section{Experimental Results}
\label{experimentalresults}

After providing details on how the model is trained in Section \ref{subsec:training}, we show the latent space visualization of the class clusters to visualize how our model classifies unlabeled data in Section \ref{subsec:Latent}. Then, we report the comparison of our proposed model GAN-CNMP with the base model CNMP in Section \ref{subsec:Comparison}. Finally, in Section \ref{subsec:collab}, we show how the humans collaborated with GAN-CNMP using the interactive paint environment we created. 

\subsection{Training}
\label{subsec:training}

Adam optimizer\cite{kingma2014adam} is used with a learning rate of 1e-4, to train the CNMP and the discriminator networks until convergence. At each optimization iteration, a random number of 2D locations and timestep tuples are selected from the sketch trajectory. The maximum number of tuples that can be selected for training is specified as 25. Each sample is fed to an encoder with 3 hidden layers where each layer has 128 hidden units. Rectified Linear Unit (RELU) activations are used between the layers. After the average 128D latent representation is obtained, a target timestep is chosen again randomly from the trajectory. This target timestep is then concatenated by the 5 preceding and 5 latter target timesteps resulting in a window size of 11. The average representation is copied 11 times to be concatenated with the corresponding 11 target timesteps. The resulting vectors are fed to the query network to obtain the mean and standard deviation of the 2D values at the 30 respective target timesteps. The query network has the same architecture as the encoder except for the input and output layers that have sizes 129 and 4 respectively. A 2-dimensional Gaussian distribution is formed for each target timestep. The network is optimized to increase the probability that the true value is sampled from that Gaussian distribution. The discriminator network consists of 2 hidden layers of size 16 followed by a hidden layer of size 4 and an output layer of 1. Similar to the generator, RELU activations are used. Binary cross-entropy loss is used to update the generator and the discriminator. The generator loss decreases and the discriminator loss increases until convergence as seen in Figure \ref{training}. The discriminator loss is initially low because the generator does not have the proficiency to generate realistic samples. As the aptitude of the generator increases the discriminator loss increases, and starts to converge. The model is optimized for 10k iterations in each epoch. For each epoch, we plot the discriminator decisions in Figure \ref{training}. The white overlays represent standard deviation. The number of times the discriminator distinguishes the fake samples as the real ones increase after 50K epochs and converge. After 200K epochs, the discriminator can separate the fake ones as real 50 \% of the time. 

\begin{figure}[h]
\vskip 0.2in
\begin{center}
\centerline{\includegraphics[width=\linewidth]{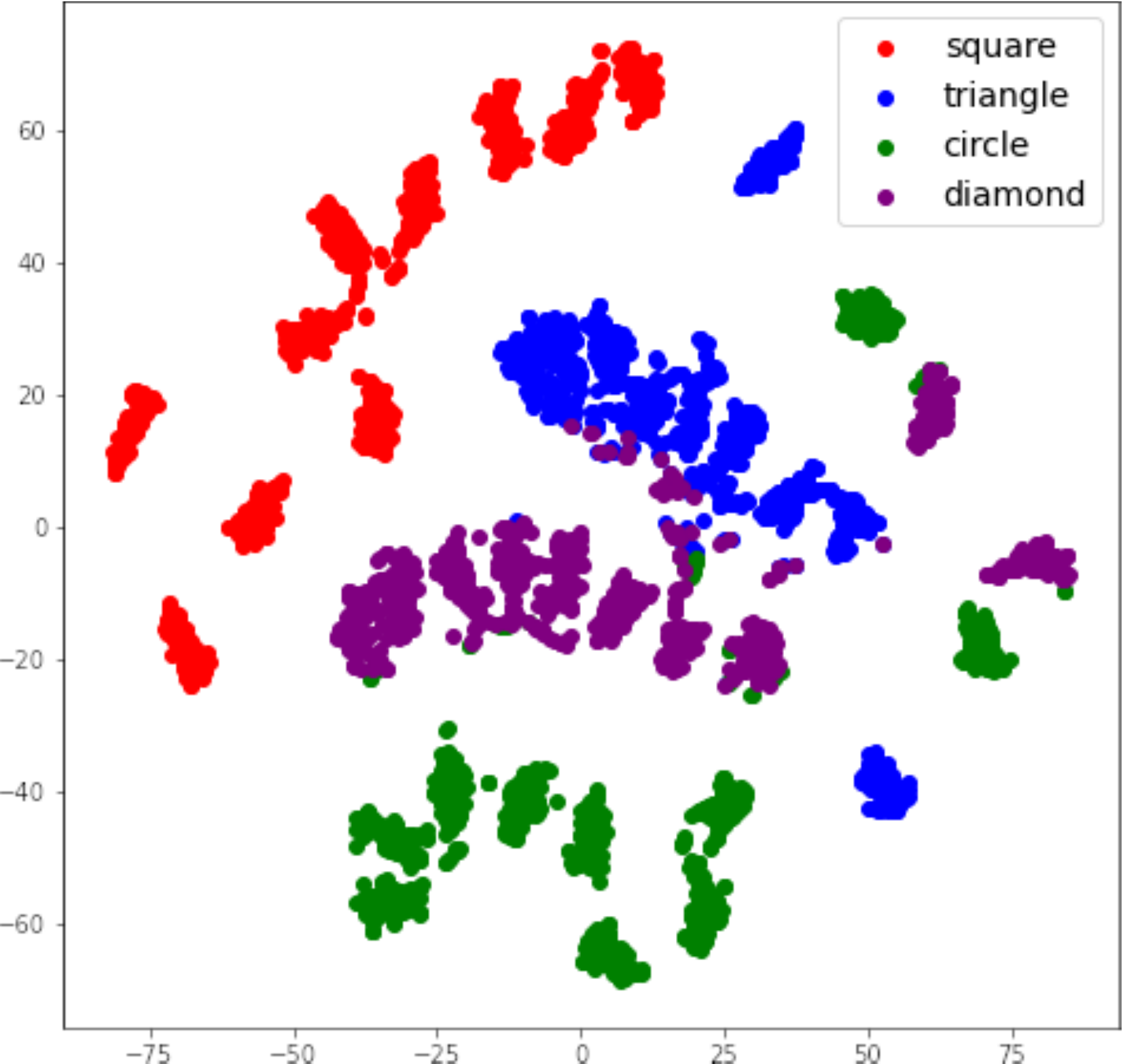}}
\caption{2D T-SNE visualization of the GAN-CNMP latent space}
\label{latent}
\end{center}
\vskip -0.2in
\end{figure}

\subsection{Latent Space Visualization}
\label{subsec:Latent}

In order to show that our GAN-CNMP model can automatically learn how to discriminate the unlabeled data in the latent space, we visualized the latent space by using T-SNE \cite{tsne} method. The latent space which has originally a size of 128 is reduced to 2 dimensions in order to plot the samples from the different shapes. The observation samples taken from 4 shapes are visualized in Figure \ref{latent}. Each colored dot represents a random observation sample from the respective shape. It can be seen that each sample is clustered within their shape type which indicates that our framework can learn and construct a distribution for each of the shapes during the training.

Although the classes are not provided during training, the model does a decent performance classifying the shapes. The visualization also shows that the square class is the best distinguished model, which is a reasonable consequence since the characteristics of the square drawings are more differential than the others. Subsequently, circle and diamond classes are inherently more challenging than the rest. This is also a reasonable challenge for our framework since the upper half of the diamond is nearly the same drawing compared to the upper side of the triangle.

\subsection{Comparison with base model}
\label{subsec:Comparison}

\begin{figure}[h]
\vskip 0.2in
\begin{center}
\centerline{\includegraphics[width=\columnwidth]{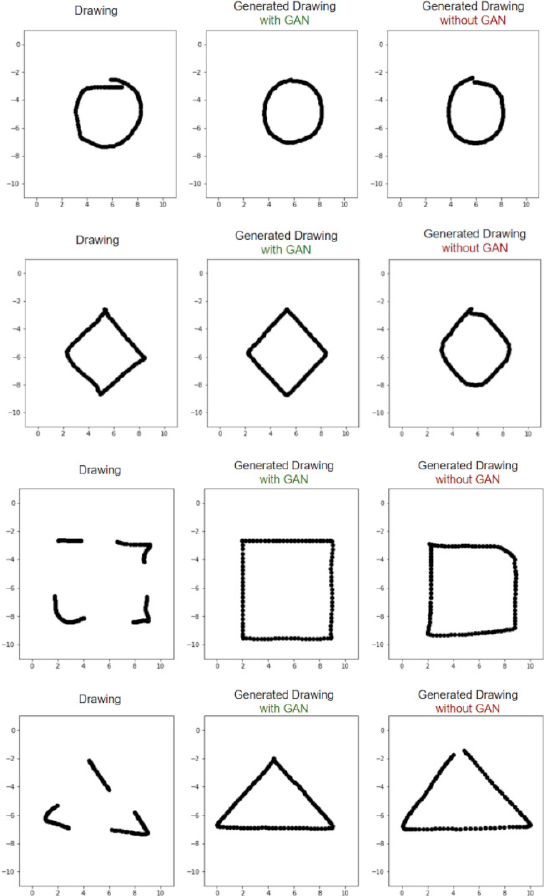}}
\caption{Comparison of the GAN-CNMP with the base model. (Left) Given user drawings. (Middle) Generated drawing by GAN-CNMP. (Right) Generated drawing by base model. The first/second row shows the examples of auto-correct feature where the third/fourth row illustrates auto-correct examples.}
\label{fig:compare}
\end{center}
\vskip -0.2in
\end{figure}

In this experiment, we will report the comparative analysis done with CNMP(base model) and GAN-CNMP in order to show the improvement of the adversarial component added to our framework. Auto-complete and auto-correct features are compared with both models which are GAN-CNMP and the base model which does not use any generative component in its model architecture. Non-perfect and incomplete shapes collected from the users from each class in the dataset are provided to both GAN-CNMP and CNMP models as seen in Figure \ref{fig:compare}. 

Figure \ref{fig:compare} shows two auto-correction examples in the first two rows. Two non-perfect drawings, circle, and diamond which are collected from the user are regenerated using both GAN-CNMP (middle) and the base model which is CNMP (right). The auto-correction examples demonstrate that the drawings generated using GAN-CNMP are more consistent with the ground truth shapes (left). More specifically, in the circle example, the upper left and top regions of the circle drawing generated by the base model are rugged and the two ends of the circle are not connected. Likewise, the diamond shape example is shown in the second row also indicates that drawing regenerated without GAN is not able to produce sharp corners and the shape is broken compared to the drawing regenerated by our proposed model GAN-CNMP which has sharp corners and straight edges.

The third and the fourth rows of Figure \ref{fig:compare} shows two auto-complete examples where incomplete and non-perfect shapes are used for visual comparison of both models. Similar to the auto-correction task, generated drawings by GAN-CNMP are smoother, more complete, and consistent. In the square experiment in the third row, the up-right corner of the square generated by the base model is bent and broken, and the location and orientation of the bottom left corner/edge are shifted unevenly. On the other GAN-CNMP is able to regenerate the square with sharp and flat corners and edges. Similarly, the triangle example that is shown in the fourth row also shows that the generated drawing by the base model is an open and broken shape compared to the perfect and closed shape generated with GAN-CNMP.

\begin{figure}[h]
\vskip 0.2in
\begin{center}
\centerline{\includegraphics[width=\columnwidth]{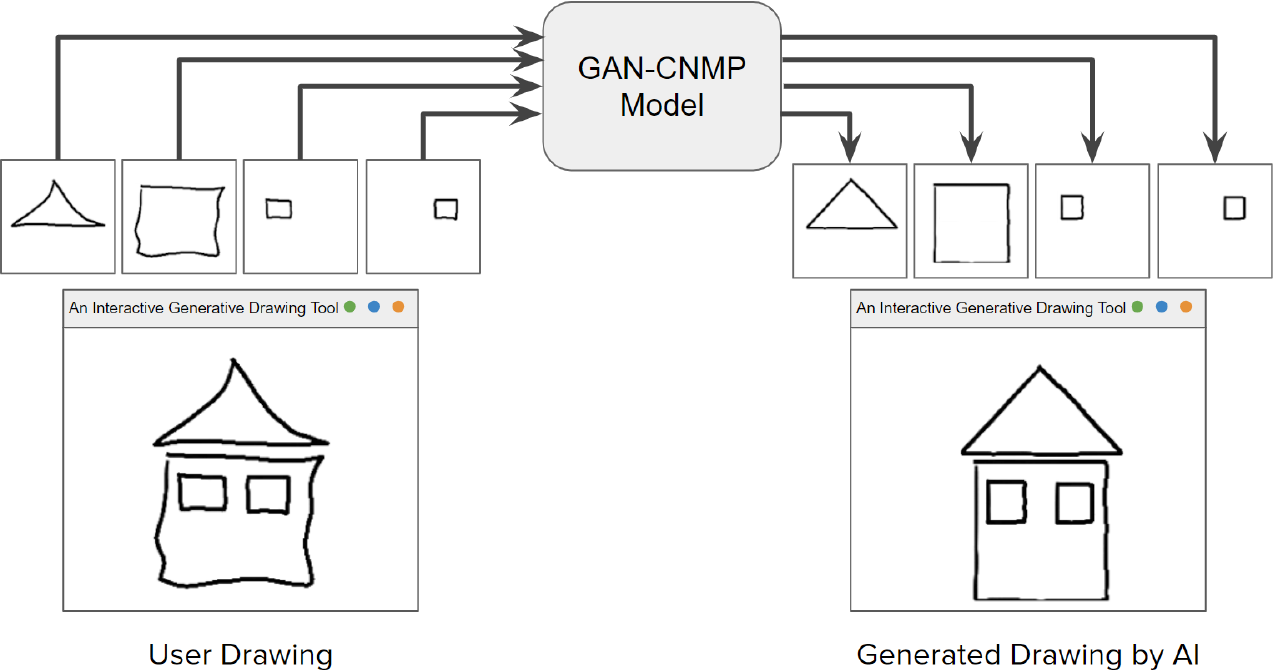}}
\caption{Pipeline of the human-AI collaboration environment}
\label{human}
\end{center}
\vskip -0.2in
\end{figure}

\subsection{Human-AI Collaboration}
\label{subsec:collab}

In this subsection, we will first explain our environment and pipeline, and then show the results of our human-AI collaboration experiment. Our aim in tackling the sketch generation task is to facilitate artistic and engineering design via a useful and enjoyable environment. Hence, we prepared an interactive paint-like environment (using canvas and mouse action recording libraries of the Python) where humans can collaborate with AI in auto-completion and auto-correction tasks and generate the drawings as they like with the help of our framework. The pipeline of the proposed environment is illustrated in Figure \ref{human}. First, a user starts to draw sketches on the screen using our environment. Each cursor movement is recorded with the related time information, and drawings are separated into simple shapes in order to be sent into our framework as input. Our framework takes these drawings as observations and regenerates the drawings from scratch while preserving the characteristics of the user input. In this visualization, a non-perfect house drawing is fed to the GAN-CNMP model that consists of a triangle and squares of different sizes. The GAN-CNMP model auto-corrects each shape independently and generates a smooth and consistent version of the shape which eventually corresponds to a house sketch. The drawings that are regenerated by our framework are sent to the drawing environment in order to allow collaboration with the user.

\begin{figure}[t]
\vskip 0.2in
\begin{center}
\centerline{\includegraphics[width=\columnwidth]{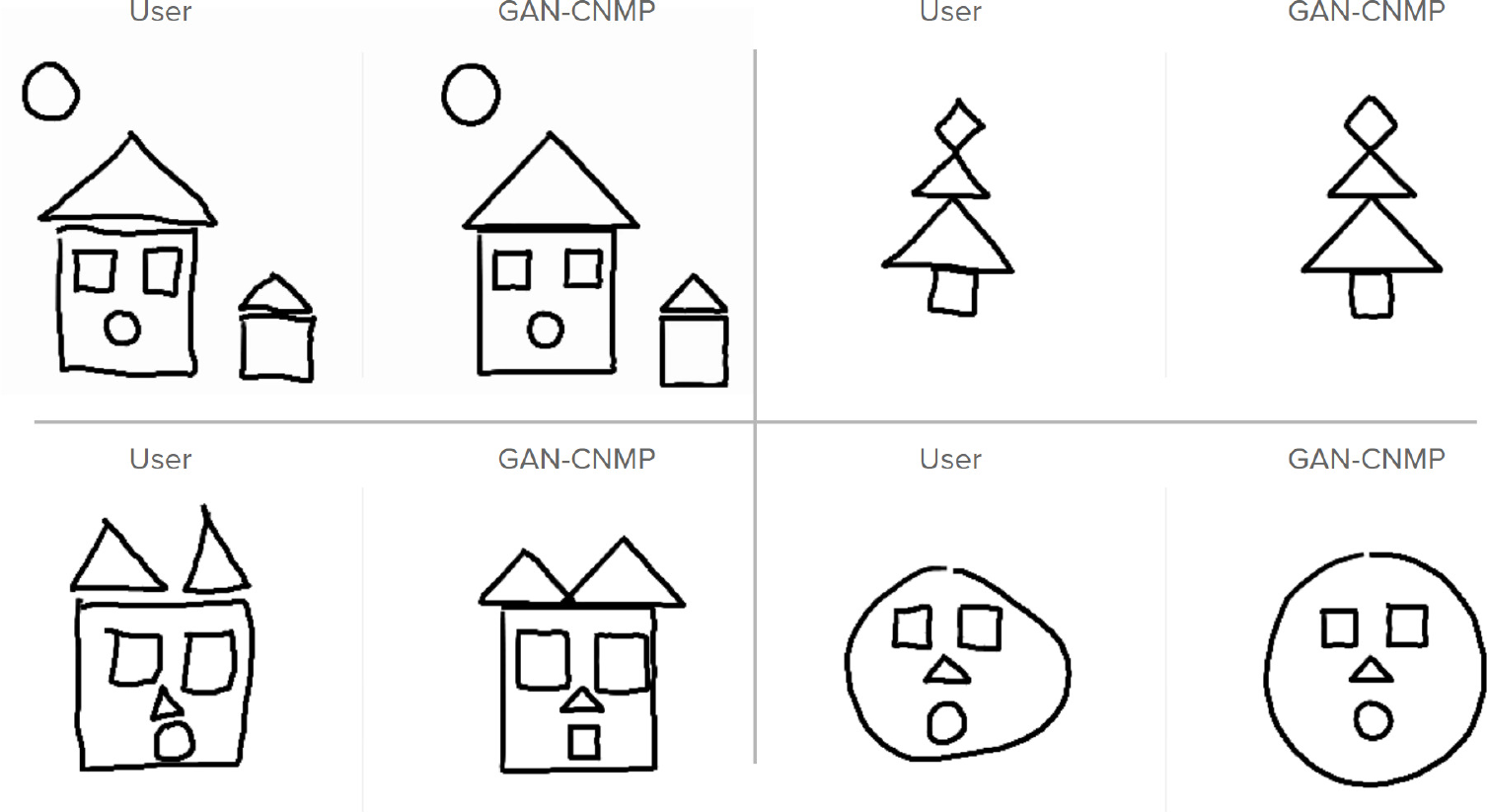}}
\caption{Example results of human-AI collaboration experiment. Experiment video : \url{https://youtu.be/Kg_Azcrao88}}
\label{collab}
\end{center}
\vskip -0.2in
\end{figure}

To test our environment with real participants, we asked some people to use our framework. At first, our aim was to collect data and feedback from as many as people we have encountered on the school campus, but because of the COVID-19 restrictions, we could only ask our families at home to participate in our experiment. First, we asked the participants to perform simple cursor movements on a blank screen in order to get them familiar with our equipment and to set the mouse DPI (dots per linear inch) to a value that they are most comfortable with. After that, we asked participants to perform any drawing that they want to draw in our environment. While doing that, we simultaneously run the pipeline that we explained in the paragraph above. The results of the four example human-AI collaboration with the participants are shown in Figure \ref{collab}. As can be seen in the examples, our framework can regenerate the drawing made by humans by auto-correcting them while preserving the characteristics of the user input. Unfortunately, the lack of participant numbers due to the COVID-19 prevented us from doing wider and detailed experiments in order to show the social and collaborative aspects of our environment in a more effective and spectacular way. 

\section{Conclusion}
\label{conclusion}
In this paper, we proposed a new framework, namely GAN-CNMP, that combines the advantages of generative models and conditional neural processes. We showed that using drawing trajectories which are functions from time to cursor positions, a generative sketch learning end-to-end model can be trained with few samples ($\sim$ 10 samples per shape type) and unlabeled data. Our model was able to construct distributions automatically in the latent space for each shape in order to distinguish drawing types from each other. We also showed that our framework can auto-correct and auto-complete the drawings made by users as well as producing better results than the base model in terms of shape consistency and smoothness. Finally, we showed that a creative environment can be constructed in order to collaborate with humans. In the future, we aim to show the advantages of our model with the other state-of-the-art sketch learning frameworks. We also aim to expand our participant number in the human-AI collaboration experiment and collect more feedback from more people.





\bibliography{example_paper}
\bibliographystyle{icml2020}

\end{document}